\documentclass[preprint,pra,aps]{revtex4-1}
\usepackage{amsmath,esint}
\usepackage{amsfonts}
\usepackage{amssymb}
\usepackage{gensymb}
\usepackage{siunitx}
\usepackage{verbatim}
\usepackage{graphicx}
\usepackage{textcomp}
\usepackage{times}
\usepackage{subcaption}
\usepackage{ragged2e}
\DeclareCaptionJustification{justified}{\justifying}
\usepackage{xcolor,colortbl}
\definecolor{LightCyan}{rgb}{0.88,1,1}
\definecolor{LightMagenta}{rgb}{1,0.88,1}
\newcolumntype{a}{>{\columncolor{LightCyan}}c}
\newcolumntype{b}{>{\columncolor{LightMagenta}}c}

\begin{document}

\title{Dirac-Maxwell correspondence: Spin-1 bosonic topological insulator for light}

\author{Todd Van Mechelen}
\author{Zubin Jacob}
\email{zjacob@purdue.edu}
\affiliation{Birck Nanotechnology Center and Purdue Quantum Center, Department of Electrical and Computer Engineering, Purdue University, West Lafayette 47907, Indiana, USA}

\begin{abstract}

Fundamental differences between fermions and bosons are revealed in their spin and distribution statistics as well as the discrete symmetries they obey (charge, parity and time). While significant progress has been made on fermionic topological phases of matter with time-reversal symmetry, the bosonic counterpart still remains elusive. We present here a spin-1 bosonic topological insulator for light by utilizing a Dirac-Maxwell correspondence. Departing from structural photonic approaches which mimic the pseudo-spin-\textonehalf{} behavior of electrons, we exploit the integer spin and discrete symmetries of the photon to formulate a distinct bosonic topological phase of matter. We introduce a bosonic Kramers theorem and the photonic equivalent of topological quantization, which arises solely from photon spin. Our continuum field theory predicts that photons acquire a mass in the presence of a spatio-temporally dispersive degenerate chirality, a unique form of magneto-electric coupling inside matter fundamentally different from well-known chirality, magneto-electricity, gyrotropy or bi-anisotropy.  We predict that this unique dispersive (non-local) degenerate chiral medium has anomalous parity and time-reversal symmetries and if found in nature will exhibit a gapped Quantum spin-1 Hall bosonic phase. Photons do not possess a conductivity transport parameter which can be quantized (unlike electronic systems), but we predict that photon spin quantization of symmetry-protected edge states is amenable to experimental isolation leading to a new bosonic phase of matter.

\end{abstract}

\maketitle 

\subparagraph*{Introduction}\label{sec:Intro}

Topological phases of electronic materials (eTI) exhibit a host of intriguing phenomena such as symmetry-protected edge states, spin-momentum locking \cite{VanMechelen2016}, quantized magneto-electric effect \cite{Qi1184}, Weyl points \cite{CTChan2016,Lu622} and Fermi arcs. The phenomena in non-interacting electronic systems can be traced back to time-reversal and parity symmetry properties \cite{Kane_2_2005} of the band structure and underlying Hamiltonian. A fundamental ingredient is the spin-\textonehalf{} of the electron \cite{Haldane1988}; enabling the definition of topological invariants such as the spin Chern number and $\mathbb{Z}_2$ invariant in the Quantum spin Hall phase \cite{Kane2005}, which can be related to experimentally observed electronic transport properties (eg: Hall conductivity \cite{Bernevig1757,Chen178}).

Foundational work in symmetry-protected topological (SPT) phases \cite{Chen1604} has revealed that bosons are another avenue for topological materials \cite{Metlitski2013,Vishwanath2013}. However, these bosonic topological phases of matter require interactions to be present \cite{Senthil2013}, which is distinct from fermions. Although the photon in vacuum is a neutral non-interacting particle (being its own antiparticle), photons can interact through matter and therefore they are the best candidate for both bosonic physics and technological applications. Thus, establishing a consistent topological field theory for the photon in a bosonic framework is absolutely critical to advancing the science of topological phases.

Recent interest in photonics has focused on mimicking topological phenomena in electronics, using photonic crystals \cite{Haldane2008} that exploit the correspondence between Schr{\"o}dinger's and Maxwell's equations. This requires a pseudo-spin-\textonehalf{} electromagnetic field \cite{Khanikaev2013,Lu2013,Slobozhanyuk2017,HafeziM.2013,Karzig2015} for systems with time-reversal symmetry and synthetic gauge fields or nonlinearity for those without \cite{Wang2009,Rechtsman2013,jin2016topological,Alu2017}. However, these systems do not take into account the intrinsic spin-1 nature of the photon, nor the fundamental difference in time-reversal \cite{Lu2014} between the photon and electron. Furthermore, the topological invariants ignore dispersion in matter and cannot be defined for continuous natural media or metamaterials \cite{Silveirinha2015,Gao2015} but necessarily rely on band structure similar to electronic crystals.

Our contribution in this paper is the foundation of bosonic topological phases of matter with spin-1 quantized edge states. We provide the first definition of topological invariants utilizing the spin-1 vector fields of the photon. We emphasize our approach marks a distinct departure from previous pseudo-spin-\textonehalf{} based works in the field of topological photonics. We achieve this by establishing a Dirac-Maxwell correspondence, a paradigm shift from existing Schr{\"o}dinger-Maxwell analogies; along with bosonic time-reversal and parity symmetry based integer spin quantum numbers. Furthermore, we show the existence of a practical topological phase employing a unique matter induced magneto-electric coupling for photons which achieves the first true bosonic counterpart of the celebrated quantum spin Hall phase of matter (QS\textsuperscript{1}HE as opposed to QS\textsuperscript{1/2}HE). Intriguingly, temporally and spatially dispersive response parameters (i.e. non-locality), commonly overlooked and considered detrimental in topological systems, are necessary features in our spin-1 bTI emerging naturally from symmetry constraints. This bosonic topological phase is fundamentally connected to anomalous properties of parity and time-reversal symmetry for bosons. Finally, we prove a bosonic Kramers theorem and discover quantized spin in symmetry-protected helical edge states which does not occur in any existing photonic crystal or metamaterial designs.

We emphasize that the magneto-electric coupling we have discovered is temporally as well as spatially dispersive, yet time-reversal symmetric, and does not utilize pseudo-fermionic operators. This is distinct from all previous work in topological photonics including non-reciprocal Tellegen media \cite{He4924}, gyrotropic media or bi-anisotropic media and does not fall into a recent classification introduced for topological photonic crystals \cite{Lein2017}. For completeness, we also mention that our predicted bosonic phase of matter leads to an effective field theory of two-component bosons (TM and TE polarization) interacting through the magneto-electric coupling \cite{Regnault_2013,Lu2012}. Hence, our work is consistent with previous results on even number quantization in interacting bosonic SPT phases \cite{Metlitski2013}. 

\subparagraph*{Dirac-Maxwell correspondence}\label{sec:DMC}

The correspondence between Dirac's and Maxwell's equations is illuminated in the Reimann-Silberstein (R-S) basis \cite{Bialynicki-Birula2013,Barnett2014}, which we utilize to develop a topological field theory of the photon. In the R-S basis $\pmb{\Psi}=(\mathbf{E}+i\mathbf{H})/\sqrt{2}$, Maxwell's equations in vacuum are naturally combined into a first-order wave problem - similar in form to the massless Dirac equation,
\begin{equation}\label{eq:SpinMomentum}
\omega\pmb{\Psi}=\mathcal{H}^{0}_{ph}\pmb{\Psi}=\mathbf{k}\cdot\mathbf{S}~\pmb{\Psi}, \qquad E\psi=H^{0}_e\psi=\mathbf{k}\cdot\pmb{\sigma}\psi.
\end{equation}
The particles are linearly dispersing $\omega=E=k$ (i.e. massless) and obey the same spin Lie algebra. However, the two wave functions are fundamentally different - $\pmb{\Psi}$ is associated with photonic (``$ph$'') vector fields and $\psi$ with electronic (``$e$'') spinor fields. $(S_j)_{ik}=i\epsilon_{ijk}$ are the antisymmetric matrices of SO(3) corresponding to the generators of spin-1 for the photon, while $\pmb{\sigma}$ are the Pauli matrices of SU(2) and represent the generators of spin-\textonehalf{} for the electron. Nevertheless, the strikingly similar form of the two foundational equations helps us associate the generators of SO(3) with photonic spin as well as define the qualitatively identical vacuum Hamiltonians $\mathcal{H}^0_{ph}$ and $H^{0}_e$.

\begin{figure}
\includegraphics[width=0.6\linewidth]{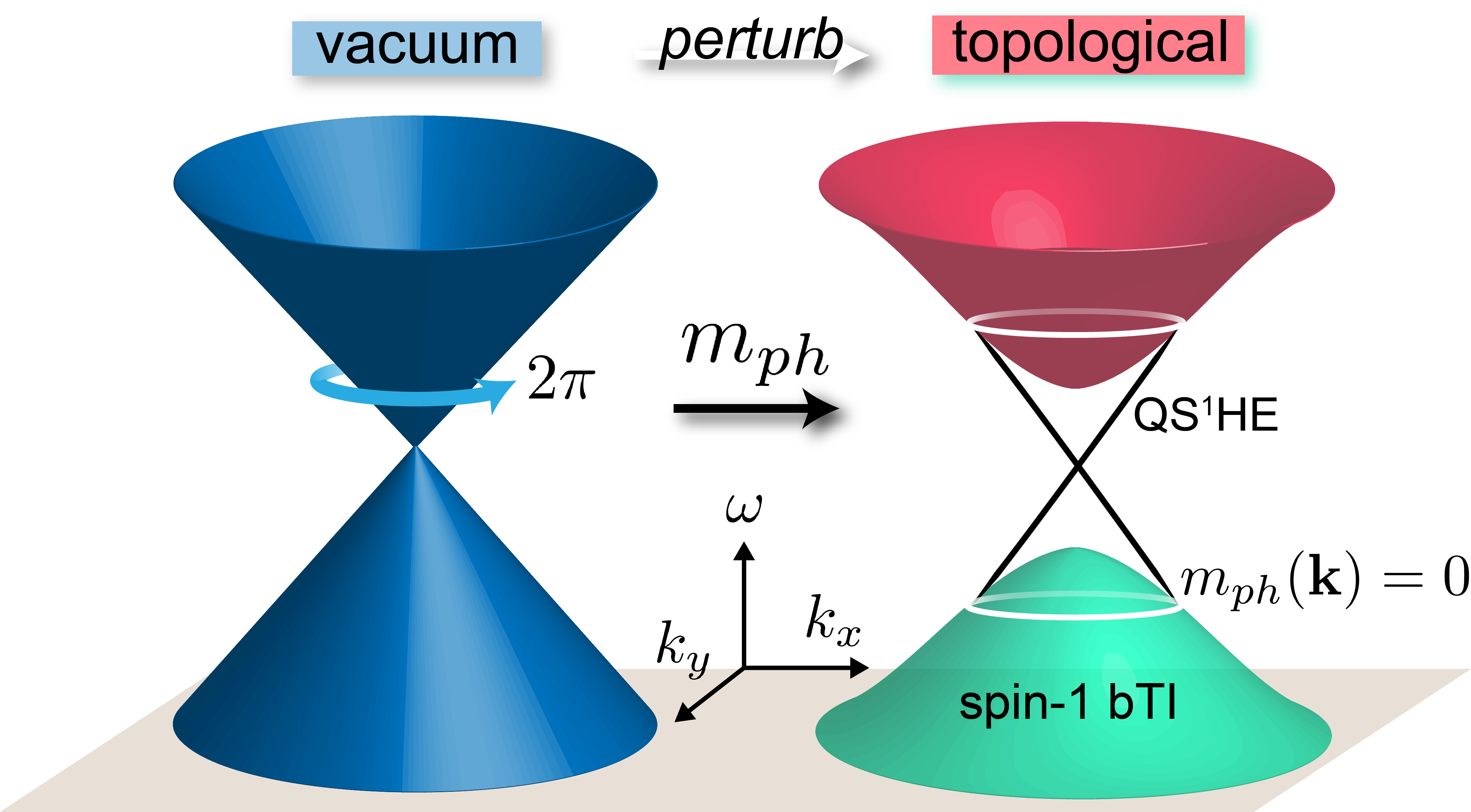}
\caption{For electrons, a bandgap is established at the zero energy point once mass is introduced. For photons, we predict that an equivalent bandgap arises from a degenerate chiral perturbation that mimics the Dirac mass. This bosonic phase of matter is realized by opening a topological bandgap with this Dirac-Maxwell photon mass $m_{ph}$. The symmetry-protected QS\textsuperscript{1}HE edge states display spin-1 quantization in striking contrast to any known phase of matter (black lines). They emerge from the transition region where $m_{ph}(\mathbf{k})=0$ passes through zero (white ring).}
\label{fig:DiracPoint}
\end{figure}

\subparagraph*{Parity-time anomaly and Bosonic Kramers theorem}\label{sec:PT}

The topological field theory of a particle is formulated by appealing to the discrete symmetries (charge, parity and time - $\mathcal{CPT}$) of the system. Interacting bTIs \cite{Chen1604,Metlitski2013,Vishwanath2013} have recently been introduced that take into account charge conservation and time-reversal symmetry. However, the photon is a fundamentally neutral particle so we must establish a different basis for the topological theory. To this end, we concern ourselves with the behavior under both time-reversal $\mathcal{T}^{-1} H(-\mathbf{k})\mathcal{T}=H(\mathbf{k})$ and parity $\mathcal{P}^{-1} H(-\mathbf{k})\mathcal{P}=H(\mathbf{k})$ symmetry. $H(\mathbf{k})$ is a Hamiltonian uniquely defined when the vacuum fields are modified by a continuous or periodic medium. First, we note the key difference in time-reversal between fermions and bosons,
\begin{equation}
\mathcal{T}_{e}^{2}=-1, \qquad \mathcal{T}_{ph}^2=+1.
\end{equation}
These operators can be rigorously defined through the Dirac-Maxwell correspondence and the non-trivial phase ($e^{i\pi}=-1$) arises precisely from the half-integer spin of the electron, clearly absent for integer spin photons. On the other hand, parity operation for both particles is trivial $\mathcal{P}_e^2=\mathcal{P}_{ph}^2=+1$. For electrons, parity and time-reversal commute $[\mathcal{P}_e,\mathcal{T}_e]=0$ revealing that a cyclic operation of $\mathcal{P}$ and $\mathcal{T}$ maintains the non-trivial phase $(\mathcal{PT})_e^2=-1$. We therefore argue that photons can acquire this phase only if parity and time-reversal anti-commute, 
\begin{equation}\label{eq:PTSymmetry}
\{\mathcal{P}_{ph},\mathcal{T}_{ph}\}=0, \qquad (\mathcal{PT})_{ph}^2=-1.
\end{equation}
We address this phenomenon as the parity-time ($\mathcal{PT}$) anomaly which is possible because $\mathcal{T}$ is an anti-linear operator. Notice that parity is odd under time-reversal (and vice versa) so the photon acquires a phase under cyclic operation of $(\mathcal{PTPT})_{ph}=(\mathcal{PT})_{ph}^2=-1$. Hence, our search for the spin-1 bosonic topological insulator (bTI) is directed at a class of photonic media exhibiting this parity-time anomaly.

The existence of the $\mathcal{PT}$ anomaly allows one to prove a bosonic equivalent of Kramers theorem (see supp. info.). Combining both symmetries, the Hamiltonian must commute $[(\mathcal{PT})_{ph},H]=0$, and this guarantees degeneracy in the states. These degenerate states $|\psi_{\mathbf{k}}^{\pm}\rangle$ have a natural basis of definite parity,
\begin{equation}\label{eq:KramerPairs}
\mathcal{P}_{ph}|\psi_{-\mathbf{k}}^{\pm}\rangle=\pm |\psi_\mathbf{k}^{\pm}\rangle, \qquad |\psi_{\mathbf{k}}^{\pm}\rangle=\mathcal{T}_{ph}|\psi_{-\mathbf{k}}^{\mp}\rangle,
\end{equation}
which are orthogonal $\langle \psi_\mathbf{k}^{\pm}|\psi_\mathbf{k}^{\mp}\rangle=0$ Kramer partners. We stress that unlike the eTI, Kramer partners for the bTI have opposite parity rather than spin.

\begin{figure}
\begin{subfigure}[t]{0.49\textwidth}
\includegraphics[width=0.9\linewidth]{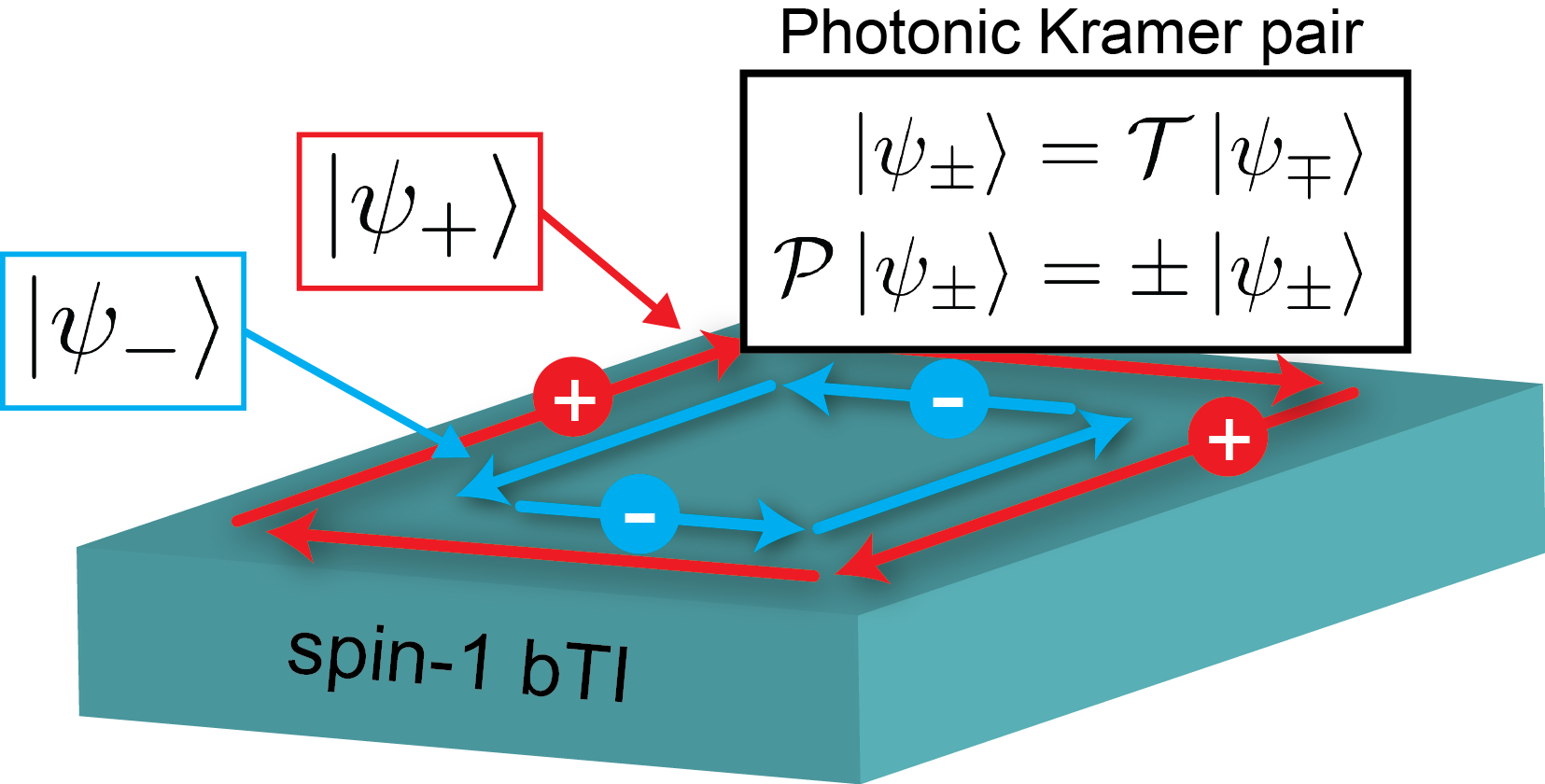}
\caption{Surface Kramer pair.}
\label{fig:SurfaceKramer}
\end{subfigure}
\begin{subfigure}[t]{0.49\textwidth}
\includegraphics[width=0.9\linewidth]{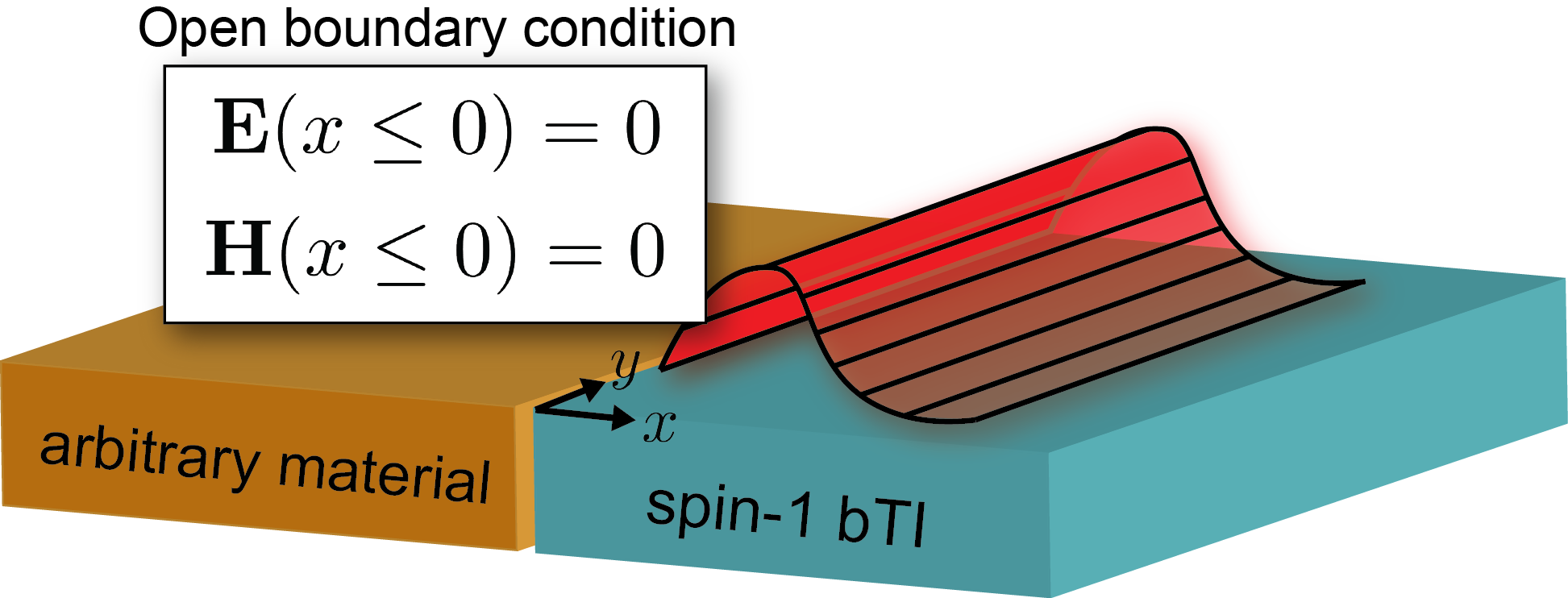}
\caption{Topological boundary condition.}
\label{fig:BoundaryCondition}
\end{subfigure}
\begin{subfigure}[t]{0.49\textwidth}
\includegraphics[width=0.9\linewidth]{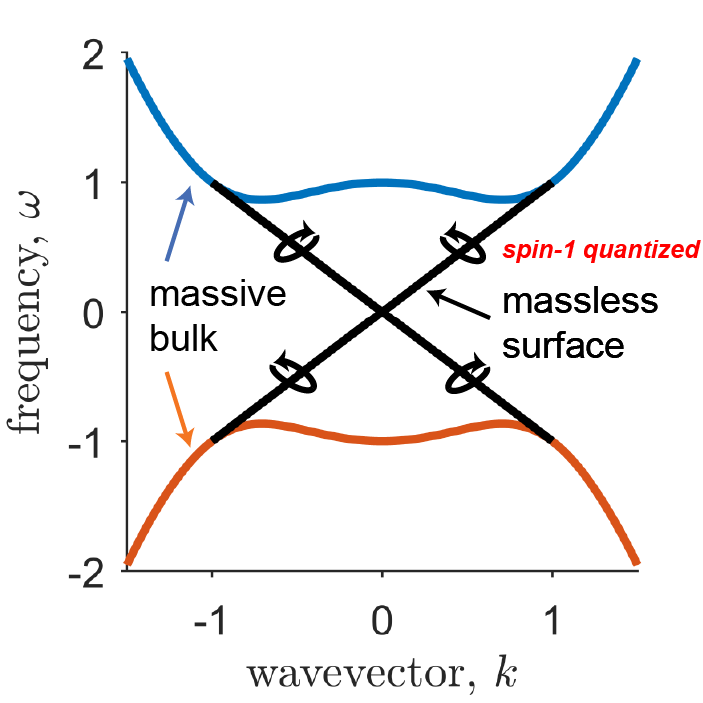}
\caption{Dispersion $\omega(\mathbf{k})$.}
\label{fig:SurfaceDispersion}
\end{subfigure}
\begin{subfigure}[t]{0.49\textwidth}
\includegraphics[width=0.9\linewidth]{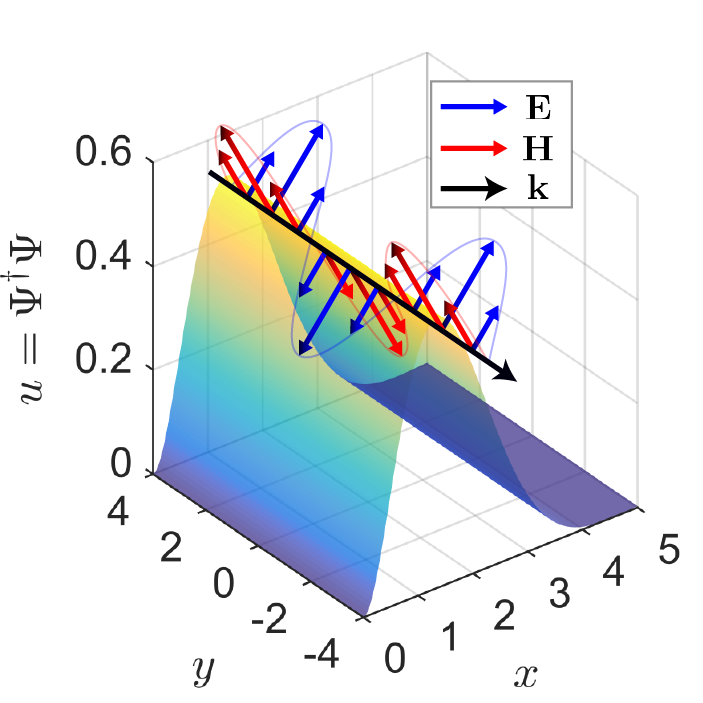}
\caption{Surface electromagnetic polarization.}
\label{fig:SurfacePolarization}
\end{subfigure}
\caption{QS\textsuperscript{1}HE edge states of the spin-1 bTI with $a=b=1$ as an example. (a) Surface Kramer pair of the bTI that counter-propagate with opposite parity. (b) Topological boundary condition that guarantees the existence of the QS\textsuperscript{1}HE at any interface with the spin-1 bTI. All components of the electromagnetic field vanish at the boundary, which is only possible due to the presence of spatial dispersion. (c) The dispersion $\omega(\mathbf{k})$ of the edge states (black) emerging from the bulk bands (blue and orange). The bulk consists of massive photons, whereas the edges support massless photons with integer (not half-integer) helical quantization. (d) Surface electromagnetic polarization at a momentum $k=0.5$. $u(x)$ is the normalized energy density and the field is confined entirely within the bTI $u(x\leq 0)=0$. The electric $\mathbf{E}$ (blue) and magnetic $\mathbf{H}$ (red) fields are $45\degree$ polarized and completely transverse to the momentum $\mathbf{k}\cdot\mathbf{E}=\mathbf{k}\cdot\mathbf{H}=0$. The counter-propagating edge states are also orthogonal, leading to robustness against backscattering.}
\label{fig:EdgeStates}
\end{figure}

\subparagraph*{Degenerate chirality and the spin-1 bTI}\label{sec:DiracMaxwell}

We now propose a 2-dimensional material and construct a spin-1 topological phase for the photon. The electromagnetic medium will exhibit a parity-time anomaly if the perturbation satisfies a $(\mathcal{PT})_{ph}^2=-1$ symmetry. From the constitutive relations, which change the symmetry properties of the field due to coupling with matter, we find that a single ingredient is required,
\begin{equation}\label{eq:ConstitutiveRelations}
\begin{bmatrix}
\mathbf{D} \\ \mathbf{B}
\end{bmatrix}=\begin{bmatrix}
\epsilon & \gamma^z S_z \\ \gamma^z S_z & \mu
\end{bmatrix}\begin{bmatrix}
\mathbf{E} \\ \mathbf{H}
\end{bmatrix},
\end{equation}
where $\gamma^z=\gamma^z(\omega,\mathbf{k})$ is a dispersive magneto-electric parameter that couples the fields antisymmetrically along $\hat{\mathbf{z}}$. Traditionally, time-reversal symmetric magneto-electric materials are associated with symmetric $\gamma=\gamma^T$ chirality. However, such media break the degeneracy of electromagnetic waves because left and right circular polarization have different refractive indices \cite{Li2013}, which cannot fulfill a $\mathcal{PT}$ anomaly. On the contrary, when the chirality is antisymmetric $\gamma=-\gamma^T=\gamma^z S_z$, degeneracy is ensured and the medium supports photonic Kramer pairs. Our theory is general but for simplicity we have assumed $\epsilon \geq 1$ and $\mu \geq 1$ are scalar constants such that the response is completely dielectric (non-metallic).

Note that particle-antiparticle symmetry for the photon ensures the electromagnetic fields are real and requires the degenerate chirality $\gamma^z(\omega,\mathbf{k})$ to exhibit temporal dispersion which is odd in frequency $\gamma^z(\omega,\mathbf{k})=-\gamma^z(-\omega,\mathbf{k}$). Simultaneously, time-reversal symmetry dictates that spatial dispersion must be even in the wavevector $\gamma^z(\omega,\mathbf{k})=\gamma^z(\omega,-\mathbf{k}$). Although this has commonly been ignored in topological photonic problems, we stress that both temporal and spatial dispersion (non-locality) is fundamental to realizing the spin-1 bTI and naturally arises from symmetry constraints.

The central result of our paper is that a spin-1 bTI emerges from degenerate chirality because it exhibits both a $\mathcal{PT}$ anomaly (see supp. info.) and opens a topological bandgap. This is exactly equivalent to the role of the electron mass in Dirac's equation (Fig.~\ref{fig:DiracPoint}). We demonstrate this explicitly using the Dirac-Maxwell correspondence, where the dynamics of the field are captured entirely by a $6\times 6$ dimensional Hamiltonian $\mathcal{H}_{ph}$. This includes both the vacuum Hamiltonian and degenerate chiral perturbation $\mathcal{H}_{ph}=v[\mathcal{H}^{0}_{ph}-\sigma_y\otimes S_z(\omega\gamma^z)]$,
\begin{equation}\label{eq:MassiveMaxwell}
\omega\Psi=\mathcal{H}_{ph}\Psi,\qquad \mathcal{H}_{ph}=v\sigma_z\otimes(k_xS_x+k_yS_y)-\sigma_y\otimes S_z( \omega\gamma^z v),
\end{equation}
where $\sqrt{\epsilon\mu}=v^{-1}$ is the apparent speed of light and $(S_j)_{ik}=i\epsilon_{ijk}$ are the antisymmetric matrices of SO(3). Here, $\Psi e^{i\mathbf{k}\cdot\mathbf{r}-i\omega t}$ is a 6-component wave function that contains all the information of the bulk electromagnetic field,
\begin{equation}\label{eq:RSBasis}
\Psi=\frac{1}{\sqrt{2}}\begin{bmatrix}
\pmb{\Psi}_+ \\ \pmb{\Psi}_-
\end{bmatrix}, \qquad \pmb{\Psi}_\pm =\frac{1}{\sqrt{2}}(\sqrt{\epsilon}\mathbf{E}\pm i\sqrt{\mu}\mathbf{H}),
\end{equation}
which are normalized in the R-S basis. Comparing with the strikingly similar massive Dirac Hamiltonian $H_e=H_e^0+\sigma_x\otimes 1_2(m_e)$, we immediately see that the photon acquires an effective mass,
\begin{equation}
m_{ph}(\omega,\mathbf{k})=-\omega\gamma^z(\omega,\textbf{k}).
\end{equation}
This photonic mass ($m_{ph}$) enters Maxwell's equations identically to the electron mass ($m_e$) of the Dirac equation,
\begin{equation}
\omega=v\sqrt{k^2+m_{ph}^2}, \qquad E=\sqrt{k^2+m_e^2}.
\end{equation}
Interestingly, $m_{ph}$ also mimics the spin-orbit interaction in graphene \cite{Kane2005} and modifies the transversality condition to $i\mathbf{k}\cdot\pmb{\Psi}_\pm=m_{ph}\hat{\mathbf{z}}\cdot\pmb{\Psi}_{\mp}$. This adds a longitudinal component to the propagating fields which is not present in any conventional photonic media.

We now utilize the parity eigenstates $\psi^\pm_\mathbf{k}$ (photonic Kramer pairs) to define the bulk electromagnetic waves for degenerate chirality - fundamentally different from the conventional $kDB$ system \cite{kong1986electromagnetic} of evaluating magneto-electric media. These are given as,
\begin{equation}\label{eq:ParityStates}
\psi_\mathbf{k}^+=\frac{1}{2}\begin{bmatrix}
v(-k\hat{\mathbf{z}}+m_{ph}\hat{\mathbf{k}})/\omega+i\hat{\pmb{\varphi}}\\
-iv(k\hat{\mathbf{z}}+m_{ph}\hat{\mathbf{k}})/\omega+\hat{\pmb{\varphi}}
\end{bmatrix}, \qquad \psi_\mathbf{k}^-=\frac{1}{2}\begin{bmatrix}
-v(k\hat{\mathbf{z}}+m_{ph}\hat{\mathbf{k}})/\omega+i\hat{\pmb{\varphi}}\\
iv(k\hat{\mathbf{z}}-m_{ph}\hat{\mathbf{k}})/\omega-\hat{\pmb{\varphi}}
\end{bmatrix},
\end{equation}
where $\hat{\pmb{\varphi}}$ is the cylindrical unit vector of $\mathbf{k}$ and $\psi^\pm_\mathbf{k}=\mathcal{T}_{ph}\psi^\mp_{-\mathbf{k}}$ are time-reversed Kramer partners. One can confirm that $\mathcal{P}_{ph}\psi_{-\mathbf{k}}^\pm=\pm\psi_\mathbf{k}^\pm$ are orthogonal eigenstates of opposite parity and reduce to vacuum when $m_{ph}=0$ (see supp. info.). We reiterate that the photonic wave functions (which are 6-component vector fields) are labeled by states of definite parity, not fixed polarization.

\subparagraph*{Spin-1 topological quantum numbers}\label{sec:TopologicalInsulators}

The distinguishing feature of our work on continuous photonic media is the robust definition of topological invariants in the presence of temporal as well as spatial dispersion, which is absent in periodic photonic crystal band structure. The stringent constraints on causal response parameters and connection to continuum topological quantum numbers is discussed in the supp. info.\cite{Silveirinha2015,Ryu2010}. To demonstrate, we take $\gamma^z(\omega,\mathbf{k})=-m_{ph}(\mathbf{k})/\omega$ which fulfills all the necessary constraints and conveniently removes any temporal dispersion in the effective mass $\partial_\omega m_{ph}=0$. From the definition of the Kramer pairs $\psi^\pm_\mathbf{k}$ (Eq.~\ref{eq:KramerPairs}), it immediately follows that,
\begin{equation}
C_\pm=\pm 2N, \qquad N=\frac{1}{4\pi}\int d^2k F_{xy}=\frac{1}{4\pi}\int^\infty_{-\infty} d^2k~\hat{\mathbf{d}}\cdot(\partial_x\hat{\mathbf{d}}\times\partial_y\hat{\mathbf{d}}),
\end{equation}
where $\mathbf{d}=k_x\hat{\mathbf{x}}+k_y\hat{\mathbf{y}}+m_{ph}\hat{\mathbf{z}}$ and $N$ is precisely the (skyrmion) winding number. For the spin-1 bTI, each skyrmion independently breaks time-reversal symmetry but combine to preserve $\mathcal{T}$, resulting in a vanishing total Chern number $C=C_+ +C_-=0$. Nevertheless, the parity Chern number $C_p=(C_+-C_-)/2$ is non-zero and quantized to an even integer $C_p=2N\in 2\mathbb{Z}$. Hence, a distinct spin-1 bosonic topological phase exists when $N\neq 0$ and essentially describes two superimposed Chern insulators.

\begin{figure}
\begin{subfigure}[t]{0.32\textwidth}
\centering
\includegraphics[width=\linewidth]{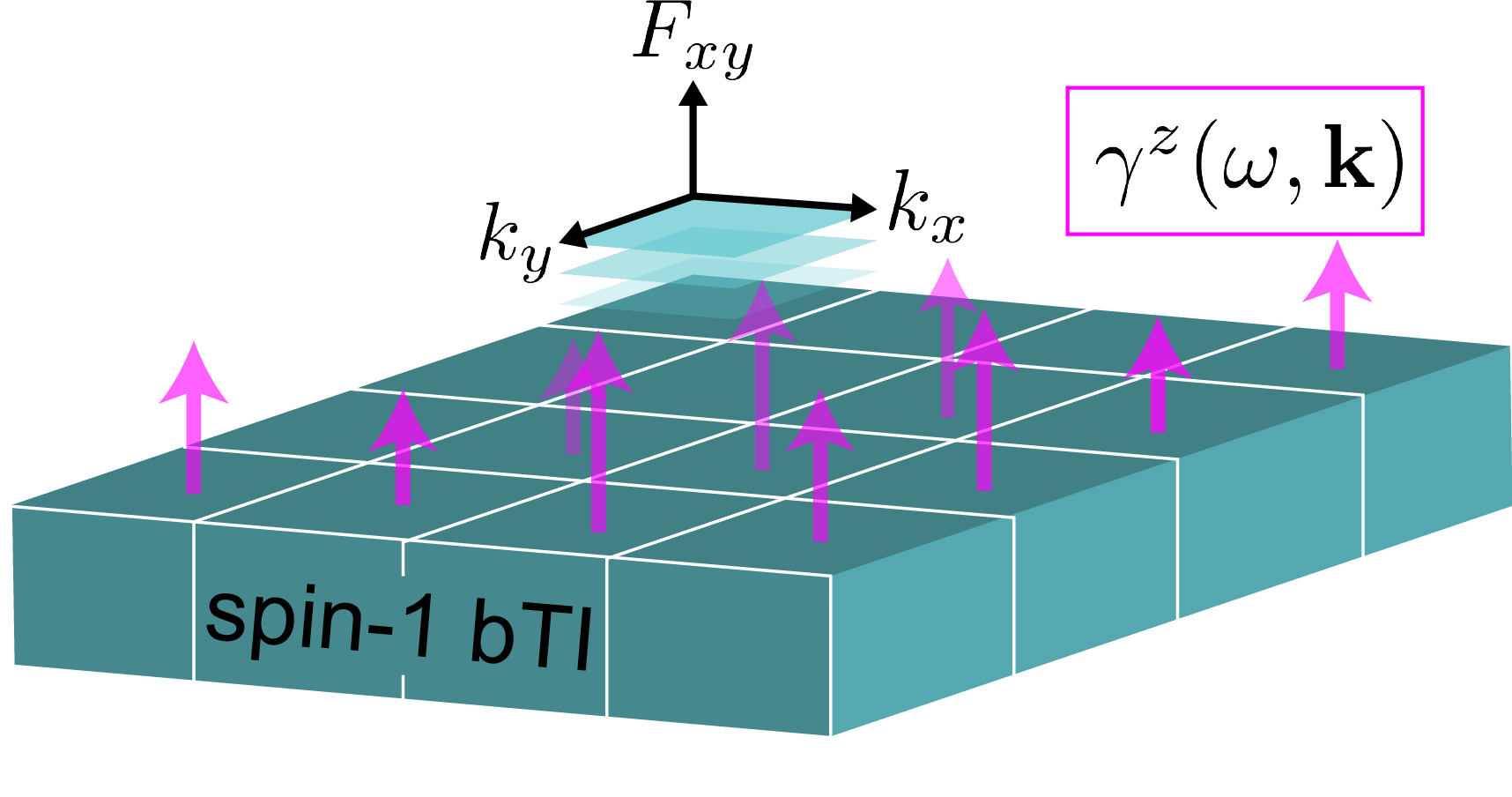}
\caption{Berry flux.}
\label{fig:BerryDiagram}
\end{subfigure}
\begin{subfigure}[t]{0.32\textwidth}
\centering
\includegraphics[width=\linewidth]{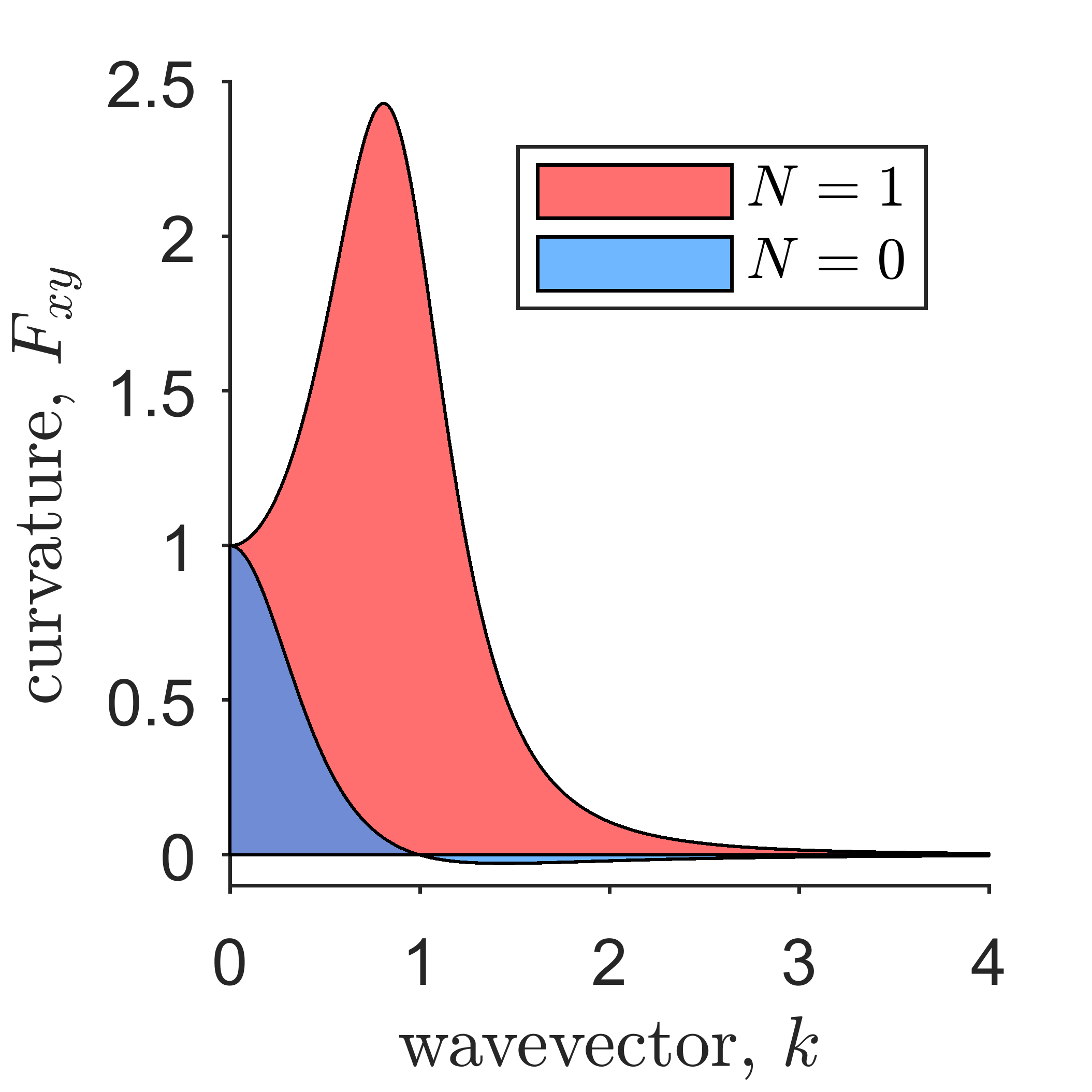}
\caption{Berry curvature $F_{xy}$. }
\label{fig:BerryCurvature}
\end{subfigure}
\begin{subfigure}[t]{0.32\textwidth}
\centering
\includegraphics[width=\linewidth]{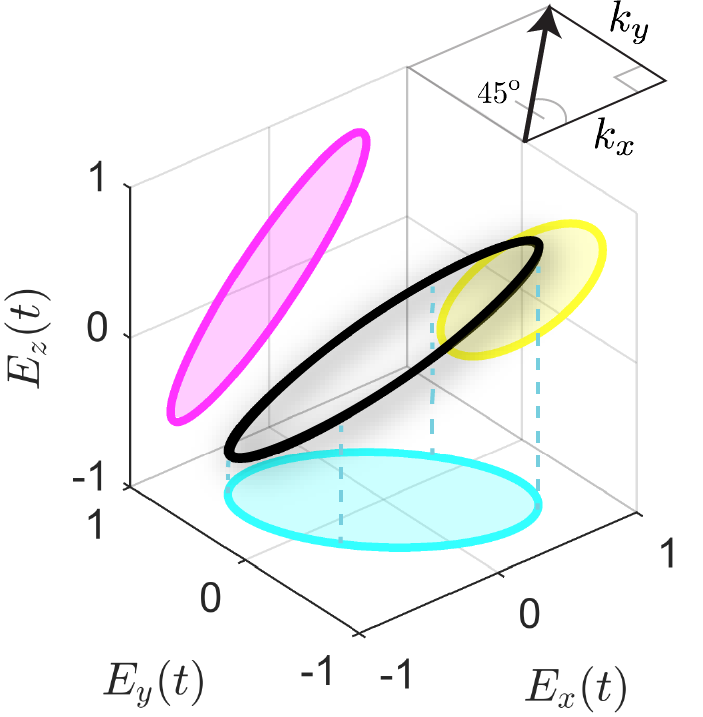}
\caption{3D electric polarization.}
\end{subfigure}
\caption{Example of a trivial and non-trivial bosonic topological phase. As a demonstration, we take the Dirac-Maxwell effective mass to be $m_{ph}(\mathbf{k})=-\omega\gamma^z(\omega,\textbf{k})=a-bk^2$, letting $a=-b=1$ for the trivial case and $a=b=1$ for the non-trivial case. Temporal and spatial dispersion emerges naturally from antiparticle and time-reversal symmetry respectively, which is fundamental to the definition of the spin-1 bTI. Here, $N=[\mathrm{sgn}(a)+\mathrm{sgn}(b)]/2$ labels the skyrmion number and $C_p=2N$ is the parity Chern invariant for each phase. (a) Berry flux through the $k_x$-$k_y$ surface. (b) Berry curvature for a photonic skyrmion. (c) Real electric field polarization $\mathbf{E}(t)$ at $k_x=k_y=1/2$ for the non-trivial phase. The electric field for a $\psi^+_\mathbf{k}$ state is generally described by a 3D polarization (black line), as opposed to the 2D polarization of conventional photonic media. The magenta, cyan and yellow lines show the elliptical projections of this 3D polarization in each of the orthogonal planes.}
\end{figure}

\subparagraph*{Quantum spin-1 Hall phase of matter}\label{sec:QSHE}

Lastly, we analyze the unique edge states of the spin-1 bTI (Fig.~\ref{fig:SurfaceKramer}), which has no counterpart in traditional surface photonics such as plasmon polaritons, Tamm states, Dyakonov or Zenneck waves \cite{Takayama2008}. These QS\textsuperscript{1}HE states are localized entirely within the bTI $\Psi(x>0)$ and propagate along the edges with intriguing open boundary conditions. Physically, the topological nature of the medium ensures all components of the electromagnetic field vanish at the edge $\Psi(x\leq 0)=0$ such that the contacting medium at $x=0$ can be completely arbitrary (Fig.~\ref{fig:BoundaryCondition}). This means the existence of the QS\textsuperscript{1}HE is guaranteed and their dispersion is immune to boundary effects. We emphasize that our spin-1 bTI differs fundamentally in this respect from previous works in photonics. Note, identical open boundary conditions are employed in topological electronics; examples range from the SSH model to the graphene spin Hall phase \cite{Delplace2011,shen2011topological}.

To uncover the edge states, we utilize a spatially dispersive form of the Dirac-Maxwell effective mass $m_{ph}=-\omega\gamma^z(\omega,\textbf{k})=a-b(k_x^2+k_y^2)$ and allow $k_x\to i\eta$ to represent the bound direction with $k_y$ along the propagating edge. Inserting into the Hamiltonian of Eq.~\ref{eq:MassiveMaxwell}, we apply the open boundary condition to discover topological edge states (not the conventional additional boundary condition of spatially dispersive media \cite{agranovich2013crystal}). We see that two counter-propagating solutions $\Psi_\pm=\Psi_\pm (x)e^{\pm ik_y y-i\omega t}$ emerge,
\begin{subequations}\label{eq:QSHEEdgeState}
\begin{equation}
\Psi_\pm (x>0)=\frac{\Psi_0}{\sqrt{2}}\begin{bmatrix}
\mathrm{sgn}(b)\mathbf{e}_\pm \\
\pm i\mathbf{e}_\mp
\end{bmatrix}(e^{-\eta_1 x}-e^{-\eta_2 x}), \qquad \omega=vk_y,
\end{equation}
\begin{equation}
\eta_{1,2} =\frac{1}{2|b|}\left[1\pm\sqrt{1+4b(bk_y^2-a)}\right], \qquad -\sqrt{\frac{a}{b}}\leq k_y\leq\sqrt{\frac{a}{b}}, 
\end{equation}
\end{subequations}
where $\mathbf{e}_\pm=(\pm i\hat{\mathbf{x}}+\hat{\mathbf{z}})/\sqrt{2}$ are the usual spin-1 helical eigenstates of the massless vacuum. Note a striking fact, the edge states touch the bulk bands at the precise points where $m_{ph}=0$ passes through zero. Evidently, the edge states only exist for the non-trivial phase when $\mathrm{sgn}(a)=\mathrm{sgn}(b)$ and are an inherent property of the bTI band structure, confirming our theory.

Applying time-reversal, the edge states are photonic Kramer pairs $\Psi_\pm= \mathcal{T}_{ph}\Psi_\mp$, which are orthogonal $\Psi_\pm^\dagger\Psi_\mp=0$ and therefore immune to backscattering. Moreover, the states are linearly dispersing $\partial_\mathbf{k}\omega=v\hat{\mathbf{k}}$ and purely transverse polarized $\mathbf{k}\cdot\mathbf{e}_\pm=0$; fundamentally distinct from conventional surface electromagnetic states that possess longitudinal fields. We must emphasize that the bulk degenerate chiral medium consists of massive photons but the edge states are massless.

Photonic transport parameters do not exhibit quantization like the electronic Hall conductivity. However, these helical edge states are spin-1 quantized along the direction of propagation, 
\begin{equation}
(\sigma_z\otimes \hat{\mathbf{k}}\cdot\mathbf{S})\Psi_\pm=\pm \Psi_\pm,
\end{equation}
with $\hat{\mathbf{k}}=\hat{\mathbf{y}}$ in this case. This also ensures the momentum is of unit magnitude $\mathbf{P}_\pm =\pm\hat{\mathbf{k}}~u_\pm$, where $\mathbf{P}_\pm$ is the Poynting vector and $u_\pm=\Psi_\pm^\dagger\Psi_\pm= \epsilon|\mathbf{E}_\pm|^2+\mu|\mathbf{H}_\pm|^2$ is the energy density. In the Reimann-Silberstein basis, Eq.~\ref{eq:QSHEEdgeState} represent states of orthogonal $45\degree$ linear polarization (not circular); $\mathbf{E}_\pm \propto \hat{\mathbf{x}}+\hat{\mathbf{z}}$ and $\mathbf{H}_\pm \propto \pm (\hat{\mathbf{x}}-\hat{\mathbf{z}})$ where the angle is determined by the sign of $\mathrm{sgn}(b)$. The dispersion and electromagnetic polarization of the QS\textsuperscript{1}HE are plotted in Fig.~\ref{fig:SurfaceDispersion} and \ref{fig:SurfacePolarization} respectively. As an aside, the spin-1 equivalent of the Jackiw-Rebbi modes \cite{Jackiw1976} are obtained by letting $m_{ph}(x)$ vary across the domain wall such that it passes through zero at $m_{ph}(0)=0$. It is shown in the supp. info. that the edge states are immune to all perturbations in $m_{ph}(x)$ like their electronic counterpart.

\begin{table}
\setlength\arrayrulewidth{1.5pt}
\centering
\caption{Dirac-Maxwell correspondence for the spin-1 bosonic topological insulator.}
\label{tab:DiracMaxwell}
\begin{tabular}{|c|a|b|}
\hline
\textbf{Property} & \textbf{Dirac} (electron) & \textbf{Maxwell} (photon) \\ \hline
2-D Hamiltonian, $H$ & $\begin{aligned} H_e &= \sigma_z\otimes(k_x\sigma_x+k_y\sigma_y)\\ &+ \sigma_x\otimes 1_2(m_e)\end{aligned}$ & $ \begin{aligned} \mathcal{H}_{ph} &= v\sigma_z\otimes(k_xS_x+k_yS_y)\\ &+ \sigma_y\otimes S_z(m_{ph} v)\end{aligned}$ \\ \hline

Dispersion relation, $\omega$ & $E=\sqrt{k^2+m_e^2}$ & $\omega=v\sqrt{k^2+m_{ph}^2}$ \\ \hline
Time-reversal operator, $\mathcal{T}$ & $\begin{aligned}\mathcal{T}_e &= 1_2\otimes\sigma_y\mathcal{K} \\ \mathcal{T}_e^2 &= -1\end{aligned}$ & $\begin{aligned}\mathcal{T}_{ph} &= \mathcal{K} \\ \mathcal{T}_{ph}^2 &= +1\end{aligned}$\\ \hline
Parity operator, $\mathcal{P}$  & $\begin{aligned}\mathcal{P}_e &= \sigma_x\otimes 1_2 \\ \mathcal{P}_e^2 &= +1\end{aligned}$ & $\begin{aligned}\mathcal{P}_{ph} &= \sigma_y\otimes 1_3 \\ \mathcal{P}_{ph}^2 &= +1\end{aligned}$\\ \hline
Commutator & $[\mathcal{P}_e,\mathcal{T}_e]=0$ & $\{\mathcal{P}_{ph},\mathcal{T}_{ph}\}=0$ \\ \hline
Parity-time operator, $\mathcal{PT}$ & $(\mathcal{PT})_e^{2}=-1$ & $(\mathcal{PT})_{ph}^{2}=-1$ \\ \hline
Spin, $s$ & $s=1/2$ & $s=1$ \\ \hline
Monopole strength, $Q_s$ & $Q_{1/2}=s=1/2$ & $Q_{1}=s=1$ \\ \hline
Topological charge, $g_s$ & $g_{1/2}=2Q_{1/2}=1$ & $g_{1}=2Q_{1}=2$ \\ \hline
Time-reversal invariant & $\mathbb{Z}_2$: $\nu=\{0,1\}$ & $2\mathbb{Z}_2$: $\varkappa=\{0,2\}$ \\ \hline
\end{tabular}
\end{table}

\subparagraph*{Conclusions}\label{sec:Conclusions}

In summary, we have utilized the spin-1 properties and unique symmetries of the photon to predict a bosonic topological insulator. Fundamentally different from structural photonic approaches which mimic pseudo-spin-\textonehalf{} behavior, our theoretical framework utilizes a Dirac-Maxwell correspondence and predicts that a parity-time anomaly is a sufficient condition for bosonic topological phases. Our theoretical framework shows that a degenerate chiral optical medium exhibits a parity-time anomaly and if found in nature will behave as a spin-1 bosonic topological insulator with symmetry-protected edge states. The characteristics of this QS\textsuperscript{1}H bosonic phase is revealed through helicity quantization in photon transport, fundamentally different from existing surface electromagnetic waves.

 \subparagraph*{Acknowledgements} This research was supported by the Defense Advanced Research Projects Agency (grant number N66001-17-1-4048) and the National Science Foundation (EFMA-1641101).

\bibliography{dirac_maxwell_ref.bib}

\end{document}